\documentclass[11pt]{article}
\usepackage{amsfonts}
\newtheorem{thm}{Theorem}
\newtheorem{prop}[thm]{Proposition}

\newtheorem{exmp}[thm]{Example}
\newcommand{\proof} {  {\it{Proof.}}    }
\newcommand {\normprod}[1]{ {\textrm{:}}{#1}{\textrm{:}} } 
\def \glt{\widetilde{gl}_{\infty}}
\def \glh{\widehat{gl}_{\infty}}
\def \D{{\mathcal D}}
\def \M{{\mathcal M}}
\renewcommand{\vec}{\mathbf}
\def \r{{\vec r}}
\def \e{{\vec e}}

\def \Cset {{\mathbb C}}
\def \Zset {{\mathbb Z}}
\def \Nset {{\mathbb N}}
\begin{document}
\title{{\LARGE\bf{Tau-functions as highest weight vectors for $W_{1+\infty}$
algebra}}}
\author{
B.~Bakalov
\thanks{E-mail: bbakalov@fmi.uni-sofia.bg}
\quad
E.~Horozov
\thanks{E-mail: horozov@fmi.uni-sofia.bg}
\quad
M.~Yakimov
\thanks{E-mail: myakimov@fmi.uni-sofia.bg}
\\ \hfill\\ \normalsize \textit{
Department of Mathematics and Informatics, }\\
\normalsize \textit{ Sofia University, 5 J. Bourchier Blvd.,
Sofia 1126, Bulgaria }     }
\date{}
\maketitle
\begin{abstract}
For each $\r = (r_{1},r_{2},\ldots ,r_{N}) \in \Cset^{N}$ we construct a
highest weight module $\M_{\r}$ of the Lie algebra $W_{1+\infty}.$ The highest
weight vectors are specific tau-functions of the $N$-th Gelfand--Dickey
hierarchy. We show  that these modules are quasifinite and we give a complete
description of the reducible ones together with a formula for the singular
vectors.
\end{abstract}
\vspace{-11cm}
\begin{flushright}
{\tt{hep-th/9510211}}
\end{flushright}
\vspace{10.5cm}
\section {Introduction}
    The remarkable connection between the infinite-dimensional Lie algebras and
the soliton equations was noticed by M.~Sato \cite{S} and further developed by
Date et al.\ in
\cite{DJKM}. In particular it was found that the important for the conformal
field theory Kac--Moody algebras and the Virasoro algebra play a substantial
role in soliton theory (see \cite{AvM2,vM,D}, etc.\ for more details). We would
only mention the work of several authors (see \cite{KS,K,Wit} and
references therein) where it was discovered that the partition function of
2D-quantum gravity is a tau-function for the KdV-hierarchy and also satisfies
the so called Virasoro constraints. This result can also be  interpreted as a
construction of a certain highest weight representation of the Virasoro
algebra. Later a whole class of representations of the Virasoro algebra in
terms of tau-functions was built in \cite{HH1,HH2}. Certain special functions
like Airy or Bessel functions and Hermite or Laguerre polynomials play an
important role in all above mentioned results.

    The present paper deals with similar questions but for the Lie algebra
$W_{1+\infty}.$ This algebra is the unique central extension of the Lie algebra
of regular differential operators on the circle \cite{KP1}.
In recent works (see e.g.\ references in \cite{FKRW,AFMO}) this algebra and its
reductions $W_{N}$ were found to play an important role in quantum field
theory. $W_{1+\infty}$ is also the algebra of the additional symmetries of the
KP tau-functions \cite{vM}. The representation theory of $W_{1+\infty}$ was
recently initiated in \cite{KR,FKRW,AFMO}, etc.\ In particular V.~Kac and
A.~Radul isolated a class of of $W_{1+\infty}$-modules and classified them.
These are graded modules with finite-dimensional level spaces, called in
\cite{KR} quasifinite.

   In contrast to the general theory we are interested in
concrete representations connected to classical special functions -- this time
-- Meijer's $G$-functions (see \cite{BE,Lu}).
Our construction uses a simple but beautiful idea of Kac and Schwarz
\cite{KS}. We recall it breafly. Each tau-function corresponds to a plane $W$
 from the Sato Grassmannian $Gr$ which can be considered as an infinite wedge
product $|W\rangle.$ Assume that an operator $A$ leaves the plane
$W$ invariant. Then under the boson-fermion correspondence $\sigma$ the image
$\tau_{W}(t) =\sigma(|W\rangle) $ is an eigenvector of $A$ in the (completed)
bosonic Fock space. We take $A$ to be $\zeta\partial_{\zeta}$ (recall that $Gr$
is built from the space of formal Laurent series in $\zeta$) and impose on $W$
to be invariant under the multiplication by $\zeta^{N}$ (hence $\tau_{W}$ is a
solution of the $N$-th reduction of KP hierarchy). These restrictions yield a
compatibility condition, satisfied by $W.$ Thus we come to the other classical
object -- Meijer's differential equation \cite{BE} (see eq.~(\ref{311}) below)
which is connected to the above modules. In the last section of the paper we
give explicit formulas for the singular vectors in these modules and point out
the embeddings among reducible ones.

    Although we consider that these representations have their own value we
have to point out that our first motivation in their construction was the
solution of the so called bispectral problem (see \cite{DG,W,G} and references
therein). Starting with the highest weight vectors of these modules we
build broad classes of solutions of any rank to this problem (see \cite{BHY}
and references therein).  But what we find
more important is that these modules provide a natural representation-theoretic
setting for many results in the bispectral problem including those of
\cite{DG}.
In this way we exhibit a completely new area of
applications of the crucial idea of Sato: the interplay between representation
theory of infinite-dimensional Lie algebras and soliton equations.
\section {Preliminaries on $W_{1+\infty}$ and Sato's  Grassmannian}
    An adequate representation-theoretic model for the Dirac sea is
the infinite wedge space $F=\oplus_{m \in \Zset}F^{(m)},$ defined as
follows \cite{KP1,KP2,KRa}.
Let $V=\oplus_{j \in \Zset} \Cset v_{j}$ be infinite-dimen\-sional
vector space with basis $v_{j}.$ Then $F^{(m)}$
for $m \in \Zset$ is the linear span of all semi-infinite monomials
    \begin{displaymath}
    v_{i_{0}} \wedge v_{i_{1}} \wedge v_{i_{2}} \wedge \ldots
    \end{displaymath}
such that $i_{0}>i_{1}> \ldots$ and $i_{k}=m-k$ for $k \gg 0.$
The free fermions can be realized as wedging and contracting operators:
    \begin{eqnarray*}
   &&\psi_{-j+\frac{1}{2}} ( v_{i_{0}} \wedge v_{i_{1}} \wedge
      \ldots ) = v_{j} \wedge v_{i_{0}} \wedge v_{i_{1}} \wedge \ldots \\
   &&\psi^{\ast}_{j-\frac{1}{2}} ( v_{j} \wedge v_{i_{0}} \wedge v_{i_{1}}
      \wedge \ldots ) = v_{i_{0}} \wedge v_{i_{1}} \wedge \ldots
    \end{eqnarray*}
The Lie algebra $gl_{\infty}$ of all $\Zset\times\Zset$ matrices, having only
a finite number of non-zero entries, can be represented in $F$ via
    \begin{displaymath}
    r(E_{ij})=\psi_{-i+\frac{1}{2}} \psi^{\ast}_{j-\frac{1}{2}}
    \end{displaymath}
(where $E_{ij}v_{k}=\delta_{jk}v_{i}).$ We shall need, however, a larger Lie
algebra $\glt$ of  $\Zset\times\Zset$ matrices with finite number of non-zero
diagonals. The above representation does not make sense for $\glt.$ It must be
regularized and leads to a representation
    \begin{equation} \label{213}
    \hat r (E_{ij}) =
    \normprod{ \psi_{-i+\frac{1}{2}}\psi^{\ast}_{j-\frac{1}{2}} }
\end{equation} of a central extension $\glh = \glt \oplus \Cset c$ with
central charge $c=1$ (as usual
    $\normprod{\psi_{\mu}\psi^{\ast}_{\nu}}$
    $= \psi_{\mu}\psi^{\ast}_{\nu}$ for $\nu>0$ and
    $= - \psi^{\ast}_{\nu}\psi_{\mu}$ for $\nu<0$ ).
Introduce free fermionic fields
    \begin{displaymath}
    \psi(z)=\sum_{j \in \Zset}\psi_{j-\frac{1}{2}}z^{-j}, \quad
    \psi^{\ast}(z)=\sum_{j \in \Zset}\psi^{\ast}_{j-\frac{1}{2}}z^{-j}
    \end{displaymath}
and ${\mathrm U(1)}$ current
    \begin{displaymath}
    J(z)=\normprod{\psi^{\ast}(z)\psi(z)}
      = \sum_{n \in \Zset}J_{n}z^{-n-1}.
    \end{displaymath}
The modes $J_{n}$ generate the Heisenberg algebra and each $F^{(m)}$ is its
irreducible representation with charge $m$ and central charge $c=1.$ This gives
an isomorphism, known as the boson-fermion correspondence (see e.g.\
\cite{KP2,KRa,Kac})
    \begin{eqnarray} \label{214,5}
    && \sigma\colon F \to B=\Cset [t_{1},t_{2},t_{3},\ldots;Q,Q^{-1}] ,  \\
    && J_{n}=\frac{\partial}{\partial t_{n}}\, ,\, J_{-n}=nt_{n}
      \textrm{ for } n>0 \, \,
    \textrm{ and } J_{0}=Q\frac{\partial}{\partial Q} .
    \end{eqnarray}
Introducing the states $|m \rangle=v_{m} \wedge v_{m-1} \wedge v_{m-2} \wedge
\ldots $ and the operator
    $H(t)=- \sum \limits_{k=1}^{\infty}t_{k}J_{k},$
we have for $| \varphi \rangle \in F$
    \begin{displaymath}
    \sigma (|\varphi\rangle) = \sum \limits_{m \in \Zset}
      \langle m|e^{H(t)}|\varphi\rangle Q^{m}.
    \end{displaymath}
Let
    \begin{displaymath}
    \phi (z)=\hat q+J_{0}\log z+\sum \limits_{n \neq 0}J_{n}\frac{z^{-n}}{-n}
    \end{displaymath}
be a scalar bosonic field with OPE
    $\phi (z_{1}) \phi (z_{2}) \sim \log (z_{1}-z_{2}),$
such that
    \begin{displaymath}
    J(z)= \partial \phi (z), \, \, \,  Q^{m}=\lim_{z \to 0}
      \normprod{e^{m \phi(z)}}|0\rangle.
    \end{displaymath}
Then the fermionic fields $\psi(z),\psi^{\ast}(z)$ act on $B$ via
    \begin{displaymath}
    \psi^{\ast}(z)=\normprod{e^{\phi(z)}} , \,\,
                    \psi(z)=\normprod{e^{-\phi(z)}}
    \end{displaymath}
(as usual $\normprod{J_{n}J_{m}} = J_{n}J_{m}$ if $m>n$ and $=J_{m}J_{n}$ if
$m<n$ , $\normprod{\hat q J_{0}} = \normprod{J_{0}\hat q} =\hat q J_{0}).$
\hfill \\
    The Lie algebra $W_{1+\infty}=\widehat{\D}$ is the unique central extension
of the Lie algebra $\D$ of complex regular differential operators on the circle
\cite{KP1,KR}. Denoting by $c$ the central element of $W_{1+\infty}$ we
introduce a basis
    \begin{displaymath}
    c, \, J^{l}_{k}=W(-z^{k+l}\partial_{z}^{l}) \, , \,  k \in \Zset ,l \geq 0
    \end{displaymath}
and another basis
    \begin{displaymath}
    c, \, L^{l}_{k}=W(-z^{k}D_{z}^{l}) \, , \, k \in \Zset ,l \geq 0
    \end{displaymath}
where $D_{z} \equiv D = z\partial_{z}.$ Here and further for $A \in \D$ we
denote by $W(A)$ the corresponding element from $W_{1+\infty}.$ The commutator
in $W_{1+\infty}$ can be written most conveniently by the generating series
\cite{KR}
    \begin{eqnarray}\label{221}
     &&\Bigl[ W(z^{k}e^{xD}),W(z^{m}e^{yD}) \Bigr] =   \hfill \\
     &&(e^{xm}-e^{yk})W(z^{k+m}e^{(x+y)D}) +
     \delta_{k,-m}\frac{e^{xm}-e^{yk}}{1-e^{x+y}} c.  \nonumber
     \end{eqnarray}
If we fix the basis $v_{j}=z^{-j}$ of $V$ we can consider any operator
$A \in \D$ as an element of $\glt.$ This gives an
embedding $\phi_{0}\colon \D \hookrightarrow \glt$ which can be extended to an
embedding \cite{KR} $\hat{\phi}_{0}\colon W_{1+\infty} \hookrightarrow \glh$
with $\hat{\phi}_{0}(c)=c.$ Using (\ref{213}) we obtain for $c=1$ a free-field
realization
    \begin{displaymath}
    W(A) = {\mathrm{Res}}_{z=0} \normprod{\psi(z) A \psi^{\ast}(z)}
    \end{displaymath}
for $A \in \D.$
Introducing the fields
    \begin{displaymath}
    J^{l}(z) = \sum_{k \in \Zset} J^{l}_{k}z^{-k-l-1}
    \end{displaymath}
we get
    \begin{displaymath}
    J^{l}(z) = \normprod{(\partial^{l} \psi^{\ast})(z) \psi(z)}
    \end{displaymath}
and we have a bosonic realization
    \begin{equation} \label{225}
    J^{l}(z) = \normprod{e^{-\phi(z)} \frac{\partial_{z}^{l+1}}{l+1}
                e^{\phi(z)}} .
    \end{equation}

Later we shall need a slight modification of this realization. Let
    $u(z) = u_{0}\log z + \sum_{n \neq 0} u_{n}\frac{z^{-n}}{-n}$
be a constant series and replace in (\ref{225}) $\phi(z)$ by
$\phi(z)+u(z).$
Considering $A \in \D$ in the basis
$v_{j} = e^{u(z)}z^{-j}$ gives an embedding $\phi_{u}\colon \D \hookrightarrow
\glt$ (i.e.\ $\phi_{u}(A) = \phi_{0}(e^{-u} A e^{u})$ ) and its extension
    $\hat{\phi}_{u}\colon W_{1+\infty} \hookrightarrow \glh$ is
    \begin{displaymath}
    (\hat r \circ \hat{\phi}_{u}) J^{l}(z) =
      \normprod{e^{-\phi(z)-u(z)} \frac{\partial_{z}^{l+1}}{l+1}
        e^{\phi(z)+u(z)}}.
    \end{displaymath}
For example when $u(z) = s \log z$ we obtain the embedding $\hat{\phi}_{s}$
 from \cite{KR}:
    \begin{equation}\label{228}
    \hat{\phi}_{s \log z} \Bigl( W(z^{k}e^{xD}) \Bigr) =
    \sum_{j \in \Zset} e^{x(s-j)} E_{j-k,j} -
    \delta_{k,0} \frac{e^{sx}-1}{e^{x}-1} c .
    \end{equation}
\hfill \\

    Now we shall briefly recall the Pl\"ucker embedding of the Sato
Grassmannian in the projectivization of the infinite wedge space \cite{KP2,SW}.
Let
    \begin{displaymath}
    \widetilde V = \Bigl\{ \sum_{k \in \Zset} a_{k} v_{k} \Big| a_{k} = 0 \,
    \textrm{for} \,       k \ll 0 \Bigr\}
    \end{displaymath}
be the space of formal series. The Sato Grassmannian $Gr$ consists of
all subspaces $W \subset \widetilde V$ which have an admissible basis
    \begin{displaymath}
   w_{k} = v_{k}+ \sum_{i>k} w_{ik}v_{k} \, , \quad  k=0,-1,-2,\ldots
    \end{displaymath}
(we consider only transversal subspaces).

Denote by $\widetilde F$ and $\widetilde B$ the formal completions of $F$ and
$B.$ Then to the plane $W \in Gr$ we associate a state $|W\rangle \in
\widetilde F^{(0)}$ as follows
    \begin{displaymath}
   |W\rangle = w_{0} \wedge w_{-1} \wedge w_{-2} \wedge \ldots
    \end{displaymath}
A change of admissible basis results in multiplication of $|W\rangle$ by a
nonzero constant. The tau-function of $W$ is the image of $|W\rangle$ under
the boson-fermion correspondence:
    \begin{displaymath}
   \tau_{W}(t) = \sigma(|W\rangle) = \langle 0|e^{H(t)}|W \rangle .
    \end{displaymath}
It is a formal power series in $t = (t_{1},t_{2},t_{3},\ldots ).$

Finally, recall that the Baker (or wave) function of $W \in Gr$ is a formal
series of the form
    \begin{displaymath}
   \psi_{W}(x,z) = e^{xz} \Bigl( 1+\sum \limits _{i=1}^{\infty} a_{i}(x)z^{-i}
                   \Bigr)
     \end{displaymath}
such that
    \begin{displaymath}
   w_{-k} = \partial_{x}^{k} \psi_{W}(x,z) |_{x=0} \, , \quad  k=0,1,2,\ldots
    \end{displaymath}
is an admissible basis of $W$ when $v_{j}=z^{-j}.$ It is expressed by the
tau-function via
    \begin{equation} \label{237}
    \psi_{W}(x,z) = e^{xz} \frac{\tau_{W}(x-[z^{-1}])} {\tau_{W}(x)}
    \end{equation}
where
   $\tau_{W}(x) = \tau_{W}(x,0,0,\ldots),$
   $[z^{-1}] = (z^{-1},z^{-2}/2,z^{-3}/3,\ldots ).$
\section{Generalized Bessel functions}
Fix $\r \in \Cset^{N}$ and let
    \begin{displaymath}
    P_{\r}(D) = (D-r_{1})(D-r_{2}) \ldots (D-r_{N}).
    \end{displaymath}
Consider the differential equation
   \begin{equation} \label{311}
   P_{\r}(D_{z}) \Phi(z) = z \Phi(z) .
   \end{equation}
After the substitution $z = \zeta^{N},$ it becomes an equation in $\zeta$ with
two singular points: regular at $\zeta = 0$ and irregular of rank 1 at $\zeta
= \infty$ (see e.g.\ \cite{Wa}). For every sector $S$ with a center at $\zeta =
\infty$ and an angle less than $2\pi,$ it has a solution with asymptotics
   \begin{eqnarray} \label{312}
   \Phi(z) \sim \Psi_{\r}(\zeta) =
    \zeta^{s(\r)} e^{N\zeta} \Bigl( 1+\sum_{i=1}^{\infty} a_{i}(\r) \zeta^{-i}
    \Bigr) \\
    {\textrm{for}} \; \;  |\zeta| \to \infty, \; \zeta \in S,
    \;  \zeta=z^{1/N}.  \nonumber
   \end{eqnarray}
The formal (divergent) series $\Psi_{\r}(\zeta)$ is uniquely determined by
eq.~(\ref{311}) and does not depend on the sector $S.$ The other solutions of
(\ref{311}) are obtained by replacing $\zeta$ by $e^{2\pi ik/N} \zeta$ $(0
\leq k \leq N-1).$ We have
    \begin{eqnarray*}
 &&s(\r) = \sum_{i=1}^{N}r_{i} - \frac{N-1}{2},\quad a_{0}(\r) = 1,
 \hfill\\
 &&a_{1}(\r) = \sum_{i<j} r_{i}r_{j} - \frac{N-1}{2N} \left(\sum
    r_{i} \right)^{2} + \frac{N^{2}-1}{24N}
 \hfill
    \end{eqnarray*}
and all $a_{i}(\r)$ are symmetric polynomials in $r_{1},\ldots ,r_{N}.$
    \begin{exmp}\label{E3}

    {\rm (i)} When $N=2,$ $\r = (\alpha /2, -\alpha /2),$ $x = 2z^{1/2},$
 eq.~{\rm (\ref{311})} becomes the classical Bessel's equation
    \begin{displaymath}
      x^{2}\partial_{x}^{2} \Phi + x\partial_{x} \Phi -
       (x^{2}+\alpha^{2})\Phi = 0
    \end{displaymath}
and the solution with asymptotics {\rm (\ref{312})} can be taken the Bessel's
function of the third kind $K_{\alpha}$ {\rm (}see e.g.\ {\rm \cite{BE})}.

    {\rm (ii)} For all $N$ we can take $\Phi(z)=G^{N0}_{0N}((-1)^{N}z|\r)$ be
the Meijer's $G$-function {\rm (}see e.g.\ {\rm \cite{BE,Lu})}. When
$r_{i}-r_{j} \not\in \Zset$ for all $i \neq j$ it can be expressed in terms of
generalized hypergeometric functions.
    \end{exmp}
We shall need some elementary properties of $\Psi_{\r}(\zeta),$ which follow
directly from eq.~(\ref{311}). They correspond to classical properties of
Meijer's $G$-function (see \cite{BE},~\S 5.3).
Here and further we denote by $\e_{i}$ the vector from $\Cset^{N}$ with 1 on
the $i$-th place and 0's elsewhere, and $\e = \frac{1}{N} \sum_{i=1}^{N}
\e_{i}.$ Then we have
    \begin{eqnarray}
    && \zeta^{s}\Psi_{\r}(\zeta) = \Psi_{\r+s\e}(\zeta) , \label{315} \\
    && (D_{z}-r_{i})\Psi_{\r}(\zeta) = \Psi_{\r+\e_{i}}(\zeta) .  \label{316}
    \end{eqnarray}
\section{Construction of highest weight modules}
 From now on we fix $s \in \Cset , \, N \in \Nset$ and consider $\r \in
\Cset^{N}$
such that $s(\r) = s.$ Recall that $z = \zeta^{N}.$ We choose a basis in $V$
    \begin{equation}\label{411}
    v_{k} = e^{N\zeta}\zeta^{s-k} \, , \quad k \in \Zset .
    \end{equation}
Then $\Psi_{\r}(\zeta)$ corresponds to the following element of $\widetilde V$
    \begin{equation}\label{412}
    w_{0} = \Psi_{\r}(\zeta) = v_{0} + \sum_{i=1}^{\infty} a_{i}(\r) v_{i} .
    \end{equation}
We define {\em{Bessel's plane\/}} \cite{F} $W_{\r} \in Gr$ to be the unique
plane, containing $\Psi_{\r}(\zeta)$ and invariant under $D_{z},$ and denote by
$\tau_{\r}$ its tau-function. By (\ref{316}) we can construct an admissible
basis $\{ w_{-k}\}_{k \geq 0}$ of $W_{\r}$ as follows. Choose arbitrary vectors
$\e^{(k)} = \sum_{i=1}^{N} k_{i} \e_{i}$ with $k_{i} \in \Zset_{\geq 0} \, , \,
\sum_{i=1}^{N} k_{i} = k$ and set
    \begin{equation}\label{413}
     w_{-k} = \Psi_{\r + \e^{(k)} }(\zeta) =
     v_{-k} + \sum_{i>0} a_{i}(\r + \e^{(k)} ) v_{i-k} .
    \end{equation}
The Baker function of the plane $W_{\r}$ is
    \begin{equation}\label{415}
    \psi_{\r}(x,z) = e^{-N\zeta} \Bigl( \zeta \Bigl(
    1+\frac{x}{N} \Bigr)\Bigr)^{-s}  \Psi_{\r} \Bigl( \zeta
    \Bigl( 1+\frac{x}{N} \Bigr)\Bigr).
     \end{equation}
By eq.~(\ref{311}) it follows that $W_{\r}$ is invariant under the operators
$z$ and $D_{z}$, i.e.\
    \begin{equation}\label{414}
    zW_{\r} \subset W_{\r} \, , \quad D_{z}W_{\r} \subset W_{\r} .
    \end{equation}
\hfill \\

The algebra $W_{1+\infty}$ is isomorphic to its subalgebra consisting of
elements of degrees divisible by $N$ \cite{FKN,vM}. The isomorphism is given
explicitly by
    \begin{eqnarray} \label{421}
   &&\pi_{N}\colon W(z^{k} e^{xD_{z}}) \mapsto W(\zeta^{Nk} e^{
             \frac{x}{N} D_{\zeta} } ) +
        \delta_{k,0} \Bigl( \frac{1}{1-e^{x/N}} - \frac{N}{1-e^{x}} \Bigr) c,
\\    &&\pi_{N}\colon c \mapsto Nc . \hfill \nonumber
    \end{eqnarray}
Combining it with (\ref{225}) we obtain a bosonic representation $\hat r_{s,N}
 = \hat r \circ \hat\phi_{N\zeta+s\log{\zeta}} \circ \pi_{N}$ of $W_{1+\infty}$
with central charge $c=N.$ It is easy to see that
    \begin{equation}\label{422}
    \hat r_{s,N}( W(z^{k}e^{xD_{z}}) ) = \hat r(z^{k}e^{xD_{z}}) + c_{k}(x)
    \end{equation}
where the action of $z^{k}e^{xD_{z}}$ is taken in the basis (\ref{411}),
$c_{k}(x)=0$ for $k>0$ and
    \begin{equation}\label{423}
    c_{0}(x) = \frac{1}{1-e^{x/N}} - \frac{N}{1-e^{x}} +
    \frac{e^{sx}-1}{1-e^{x}} .
    \end{equation}
Note that when $k \neq 0$ in (\ref{422}) $\hat r$ can be replaced by $r.$
 From now on we shall consider only this representation ($s,N$ being fixed) and
skip the symbol $\hat r_{s,N}.$
Denote by $\M_{\r}$ be the module generated by $\tau_{\r}.$
\hfill \\
  \begin{thm}\label{Th2}
  {\rm (i)} $\tau_{\r}$ is a highest weight vector with highest weight
  $\lambda_{\r}$, i.e.\
    \begin{equation}\label{431}
    L^{l}_{k} \tau_{\r} = 0  \; \textrm{for} \; k>0, \; L^{l}_{0}
    \tau_{\r} = \lambda_{\r}(L^{l}_{0}) \tau_{\r}.
    \end{equation}
  Here $\lambda_{\r}(L^{l}_{0})$ are certain symmetric polynomials in
  $r_{1},\ldots ,r_{N},$ for example
    $ \lambda_{\r}(L^{0}_{0}) = \sum_{i=1}^{N} r_{i},$
    $ \lambda_{\r}(L^{1}_{0}) = \frac{1}{2} \sum_{i=1}^{N} r_{i}^{2} -
     \frac{1}{2} \sum_{i=1}^{N} r_{i} .$
\hfill \\

  {\rm (ii)} The module $\M_{\r}$ is quasifinite with characteristic
polynomials
    \begin{displaymath}
    P_{\r,k}(D) = P_{\r}(D) P_{\r}(D-1) \ldots P_{\r}(D-k+1)
    \end{displaymath}
$(k \in \Nset),$ i.e.\ we have
    \begin{displaymath}
    W( z^{-k} P_{\r,k}(D_{z}) ) \tau_{\r} = 0
    \end{displaymath}
and $P_{\r,k}$ are polynomials of minimal degree with this property.
   \end{thm}
   \proof
  By (\ref{311}) and (\ref{414}) $W_{\r}$ is invariant under the operators
$z^{k}D_{z}^{l}$ for $k,l \geq 0$ and $z^{-k} P_{\r,k}(D_{z}) = (z^{-1}
P_{\r}(D_{z}) )^{k}$ for $k \geq 1.$ We use the
result of \cite{KS} that for $A \in \glt , W \in Gr$
    \begin{displaymath}
   AW \subset W \textrm{ iff } \hat r (A) \tau_{W} = const.\tau_{W}
    \end{displaymath}
and the fact that
    \begin{displaymath}
   W( z^{k}f(D_{z}) ) = \frac{1}{k} \Bigl[ W( D_{z} ),W( z^{k}f(D_{z}) ) \Bigr]
    \end{displaymath}
for $k \neq 0.$
The polynomials $\lambda_{\r}(L^{l}_{0})$ can be computed comparing the
coefficient of $|0\rangle$ in both sides of (\ref{431}) and using (\ref{413})
and (\ref{423}).
   \begin{exmp}\label{E4}
For $N=2$ and $\r = (\alpha /2, -\alpha /2)$ the Virasoro modules
$M^{\infty}_{\alpha}$ introduced in {\rm \cite{HH2,F}} are reductions of the
modules $\M_{\r}$ {\rm (}obtained by putting $t_{2}=t_{4}=t_{6}=\ldots=0).$
   \end{exmp}
To describe $\lambda_{\r}$ introduce, following Kac and Radul \cite{KR}, the
generating series
    \begin{displaymath}
    \Delta_{\lambda_{\r}}(x) = \lambda_{\r}(W(-e^{xD_{z}})) .
    \end{displaymath}
It is proved in \cite{KR} that for a quasifinite representation with first
characteristic polynomial $P_{\r}(D)$ the function
    \begin{displaymath}
   F(x) = (e^{x}-1)\Delta_{\lambda_{\r}}(x) + c
    \end{displaymath}
satisfies the equation
    \begin{displaymath}
   P_{\r}(\partial_{x}) F(x) = 0 .
    \end{displaymath}
When all $r_{i}$ are distinct, it follows that $F(x)=\sum_{i=1}^{N}
a_{i}e^{r_{i}x}.$ Because of the symmetry $a_{1}=\ldots =a_{N}.$ Using
the value of $\lambda_{\r}(L^{0}_{0})$ given by Theorem \ref{Th2}
 we get $a_{1}=\ldots =a_{N}=1.$ But
$\lambda_{\r}(L^{l}_{0})$ are polynomials in $\r$, thus we have proved
  \begin{thm}
  The generating series $\Delta_{\lambda_{\r}}(x)$ is given by
    \begin{displaymath}
   \Delta_{\lambda_{\r}}(x) = \sum_{i=1}^{N} \frac{e^{r_{i}x}-1}{e^{x}-1}.
    \end{displaymath}
  \end{thm}

The irreducible module with such generating series is called in
\cite{FKRW} a primitive module. However, our modules are irreducible only when
$r_{i}-r_{j} \not\in \Zset$ for $i \neq j$ \cite{KR}.
\hfill \\

We shall show that $\tau_{\r}$ is characterized among all formal power series
by the constraints (\ref{431}).
  \begin{prop}\label{Prop45}
  There exists only one {\rm (}up to a constant{\rm )} formal power series
$\tau$ in $t_{1},t_{2},\ldots$ satisfying
    \begin{equation}\label{451}
    J^{l}_{k} \tau = 0, \quad      J^{l}_{0} \tau = c_{l}\tau
    \end{equation}
for $0<k, \; 0 \leq l \leq N-1$ and some constants $c_{l}.$
   \end{prop}
   \proof
(cf.\ \cite{AvM,HH2}) After taking derivatives of (\ref{451}) and
letting
$t_{1}=t_{2}=\ldots =0$ one sees inductively that all derivatives of $\tau$ at
$t=0$ vanish, hence it is determined by $\tau(0).$
\section{Embeddings of the modules $\M_{\r}$ }
  \begin{thm}\label{Th52}
   Let $\r \in \Cset^{N}, \; r_{1}-r_{2} = \alpha \in \Zset_{\geq 0}.$ Then
$\tau_{\r + \e_{1} - \e_{2} }$ is a singular vector in the module $\M_{\r}$ and
is given by the formula
    \begin{equation}\label{521}
    \tau_{\r + \e_{1} - \e_{2} } = W(A)\tau_{\r} + const.\tau_{\r}
    \end{equation}
where
    \begin{equation}\label{521A}
     A =  -(\alpha+1)\frac{z^{-1}P_{\r}(D_{z}) }{ D_{z}-r_{2}}
       \Bigl( z^{-1}P_{\r}(D_{z}) \Bigr)^{\alpha}
    \end{equation}
Therefore $\M_{\r+\e_{1}-\e_{2} } \hookrightarrow \M_{\r}.$
   \end{thm}
   \proof
Because
    $L^{1}_{0}|_{t_{1}=x,t_{2}=t_{3}=\ldots =0} = \-
    \frac{x+N}{N}\partial_{x} + \- \frac{s(s-1)}{2N} + \- \frac{1-N^{2}}{12N}$
, (\ref{431}) implies
    \begin{displaymath}
    \tau_{\r}(x) = \Bigl( 1+\frac{x}{N} \Bigr)^{ \frac{N}{2} \sum r_{i}^{2} +
                    const }
     \end{displaymath}
(the inessential constant depends on $s$ and $N$ but not on $\r).$
For $\lambda \in \Cset$ we consider the sum $\tau_{\lambda}(t) = \tau_{\r}(t) +
\lambda\tau_{\r + \e_{1} - \e_{2} }(t) $ (cf.\ \cite{F}).
Then (\ref{413}) implies that
    \begin{displaymath}
     \tau_{\lambda}(t) = \sigma( (\Psi_{\r} + \lambda\Psi_{\r+\e_{1}-e_{2}})
     \wedge \Psi_{\r+\e_{1}} \wedge \Psi_{\r+\e_{1}+\e^{(1)}} \wedge \ldots )
    \end{displaymath}
is a tau-function. By (\ref{237}) its Baker function
$\psi_{\lambda}(x,\zeta)$ is equal to
     \begin{eqnarray}
      \frac{    \tau_{\r}(x) \psi_{\r}(x,\zeta) +
        \lambda\tau_{\r+\e_{1}-\e_{2}}(x) \psi_{\r+\e_{1}-\e_{2}}(x,\zeta)    }
        {    \tau_{\r}(x) + \lambda\tau_{\r+\e_{1}-\e_{2}}(x)       }
     \nonumber \\
     = \frac{     \psi_{\r}(x,\zeta) + \lambda (1+\frac{x}{N})^{N(\alpha+1)}
              \psi_{\r+\e_{1}-\e_{2}}(x,\zeta)         }
            {    1 + \lambda (1+\frac{x}{N})^{N(\alpha+1)}    }    \nonumber
    \end{eqnarray}
Using (\ref{415},\ref{315},\ref{316}) and the eq.~(\ref{311}) it is easy
to see that
    \begin{displaymath}
   \Bigl( 1+\frac{x}{N} \Bigr)^{N(\alpha+1)} \psi_{\r+\e_{1}-\e_{2}}(x,\zeta)
   = e^{-N\zeta} \zeta^{-s} A(z,D_{z}) \zeta^{s} e^{N\zeta} \psi_{\r}(x,\zeta).
    \end{displaymath}
 Therefore in the basis (\ref{411})
    \begin{displaymath}
    \tau_{\lambda}(t) = \sigma( (1+\lambda A)\Psi_{\r} \wedge
    (1+\lambda A)\Psi_{\r+\e^{(1)} }
    \wedge (1+\lambda A)\Psi_{\r+\e^{(2)} } \wedge \ldots )
    \end{displaymath}
and comparing the coefficient of $\lambda$ gives
    \begin{displaymath}
    \tau_{\r + \e_{1} - \e_{2} }(t) = r(A) \tau_{\r}(t)
    \end{displaymath}
The formula (\ref{521}) now follows from (\ref{422}). \\

Theorem \ref{Th52} gives in practice all possible embeddings among the modules
$\M_{\r}.$

\hfill \\
{\flushleft\bf\large{ Acknowledgements \\}}
\hfill \\
We thank Prof. I.~T.~Todorov for useful discussions and for his interest in
this work. This paper is partially supported by Grant MM--402/94 of Bulgarian
Ministry of Education, Science and Technologies.
\begin{small}
\renewcommand{\refname}{ {\flushleft\bf\large{ References }} }
    
\end{small}
 \end{document}